\documentclass[%
reprint,
superscriptaddress,
showpacs,preprintnumbers,
showkeys,
 amsmath,amssymb,
 aps,
]{revtex4-1}

\usepackage{graphicx}% Include figure files
\usepackage{dcolumn}% Align table columns on decimal point
\usepackage{bm}% bold math
\def\be{\begin{equation}}
\def\ee{\end{equation}}
\def\bea{\begin{eqnarray}}
\def\eea{\end{eqnarray}}
\newcommand{\ket}[1]{\mbox{$|#1\rangle$}}

\begin{document}

\title { The essence of  microphysical entities}

\author{Fang-Yu Hong}
%\email[Email address:]{honghfy@163.com}
%\email[Tel:]{86-571-86843468}
\affiliation{Department of Physics, Center for Optoelectronics Materials and Devices, Zhejiang Sci-Tech University,  Hangzhou, Zhejiang 310018, China}
%\author{Huiqin Qian}
%\affiliation{Department of Physics, Center for Optoelectronics Materials and Devices, Zhejiang Sci-Tech University,  Hangzhou, Zhejiang 310018, China}
%\author{Shi-Jie Xiong}
%\affiliation{National Laboratory of Solid State Microstructures and
%Department of Physics, Nanjing University, Nanjing 210093, China}
%\author{Jing-Li Fu}
%\affiliation{Department of Physics, Center for Optoelectronics Materials and Devices, Zhejiang Sci-Tech University,  Hangzhou, Zhejiang 310018, China}
%\author{Yang Xiang}
%\affiliation{School of Physics and Electronics, Henan University, Kaifeng, Henan 475004, China}
%\author{W.H. Tang}
%\affiliation{Department of Physics, Center for Optoelectronics Materials and Devices, Zhejiang Sci-Tech University,  Hangzhou, Zhejiang 310018, China}
%\author{Zhi-Yan Zhu}
%\affiliation{Department of Physics, Center for Optoelectronics Materials and Devices, Zhejiang Sci-Tech University,  Hangzhou, Zhejiang 310018, China}
%\author{Li-zhen Jiang}
%\affiliation{College of Information and  Electronic Engineering, Zhejiang Gongshang University, Hangzhou, Zhejiang 310018,China}
%\author{Liang-neng Wu}
%\affiliation{College of Science, China Jiliang University, Hangzhou, Zhejiang 310018, China}
\date{\today}
\begin{abstract}
 In spite of its outstanding  success, quantum mechanics remains mysterious, many problems such as wave/particle dualism and quantum nonlocality  remain open. Because a particle, e.g. a photon, is a quantum  of a corresponding quantum field,  an  arbitrary  particle state directly corresponds to a quantum field, which shows the quantum field is nonlocal.  A microphysical entity (ME) can  be taken as the corresponding quantum field which is in the state of a quantum, where the quantum field  is responsible for the ME's wave-like nature and the quantum for the ME's particle-like nature.  A quantum state directly corresponds to a quantum field.  Base on this simple model,   many big problems in quantum physics, such as the wave/particle dualism, the  collapse of the quantum state on measurement, the nonlocality in quantum entanglement, quantum teleportation, quantum swapping,   and the paradox in the delayed-choice entanglement swapping, can be solved in very simple and  natural manners.
\end{abstract}

\pacs{03.67.Lx, 03.65.Vf, 74.45.+c, 85.25.-j}
%\keywords{}

\maketitle
%\section{INTRODUCTION}
%{\it Introduction.}---

 Quantum theory (QT) has achieved unprecedented success since its invention in the early 20th century. It can precisely describe the structure and interactions of atoms, nuclei, and subnuclear particles, predict the properties of the different materials, and has led to many fundamental technological discoveries like the transistor which is the building block of all the modern electronics and computers.  On the other hand much debate concerning quantum theory has persisted to this day since its first formation. A microphysical sometimes behaves like a particle and sometimes like a wave, what really is a microphysical entity? How can we understand the wave/particle dualism.  Does the quantum state corresponds directly to a physical state? If so, then how does the collapse of the quantum state on measurement take place?  How does the measurement on one of a pair entangled particles influences the other space-like one? How can quantum teleportation and quantum swapping can be accomplished on the particles which have no direct interactions among them?  How can we get out of the paradoxical situation taking place in the delayed-choice entanglement swapping where future actions seem to influence the past already irrevocably measured events? These questions remain open today. In this respect great physicist Feynman, who presented a new formulation of quantum mechanics, at one time said  that nobody understands quantum mechanics. The fundamental importance of   searching interpretative models for the QT can be seen from the fact that the great development of QT would have been impossible without the  indefatigable tries made by Einstein (EPR), Schr\"{o}dinger (S-cat) or Bohm( HV theory) at  `reasonably'  interpreting the QT.

Here we show that a microphysical entity is really the quantum field which is in the excited state of one quantum. In this model a ME has always wave-like and particle-like properties at the same time. However, what measured in a single measurement on a ME is always its particle-like properties, and its wave-like properties are manifested in the statistical distribution of the data taken in many repeated measurements.  Being the quantum field's quantum  an arbitrary state of the particle has an instantaneous corresponding quantum field state. A quantum state is real and corresponds directly to a quantum field state.  The collapse of the quantum state on measurement is only the phenomenon where the quantum field flips to the state corresponding to the results of the measuring on the ME. It is  with the help of  the quantum field, which is not localized and extends to the whole space, that a pair of entangled particles, no matter how far the distance between them is, show nonlocal relations,  and that  quantum teleportation and quantum swap become possible. With this model the paradox in delayed-choice entanglement swapping and other delay-choice experiments  can be neatly solved too.

 When a quantum field is in its basis state, i.e.,  has no excitations (quanta), the quantum field displays itself in the form of quantum vacuum. The quantum vacuum is under no circumstances  a simple empty space; it  is a real existence and  is taken as the ``Dirac sea" where  the states of negative energy are completely occupied with electrons \cite{wgre}. Dirac predicted that a photon of energy larger than 2 times electron's rest energy $m_0c^2$ can excite an electron of negative energy into a state of positive energy  leaving a hole in the quantum vacuum which is later interpreted as a positron. Soon Dirac's prediction was experimentally confirmed in all points. Particles can be created from the quantum vacuum by external fields \cite{miri}, the experimental observation of the effects of vacuum polarization \cite{ildk, schw}, the Casimir effect \cite{skla, mkrg,mbum}, and the quantum  vacuum's influence on an atom's spontaneous emission \cite{kvah}: they all  verify the reality of the quantum vacuum.

 Next we discuss the situation where a quantum field is in the excited state of one quantum such as an electron.  Considering that the quantum vacuum with some change still exists in space the quantum field now is consisted of the quantum and the corresponding quantum vacuum.  The quantum field state $\Psi(x)$ is closely related to its excitations in the form ($\hbar=1$) \cite{msre}:
\bea\label{eq1}
\Psi(x)&=&\sum_{s=\pm}\int \frac{d^3p'}{(2\pi)^22\omega }\left[b_s({\bf p}')u_s({\bf p}')e^{ip'x}\right. \notag \\&+& \left.d_s^\dagger({\bf p}')v_s({\bf p}')e^{-ip'x}\right],
\eea
 and the state of a field excitation has a corresponding quantum field state, e.g.,  the creation operator $b_s^\dagger({\bf p})$ for an electron of mass $m$, momentum  ${\bf p}$, and spin $s$ can be expressed in terms of the quantum field $\Psi(x)$
\be\label{eq2}
b_s^\dagger({\bf p})=\int d^3xe^{ipx} \overline{\Psi}(x)\gamma^0u_s({\bf p}),
\ee
where  $\overline{\Psi}(x)=\Psi^\dagger(x)\gamma^0$, $b_s({\bf p})$ and $d_s^\dagger({\bf p})$ are the annihilation operator and the creation operator for an electron and  a positron, respectively, $\omega=\sqrt{{\bf p}'^2+m^2}$,  four-component spinor $u_s({\bf p})$ and $u_s({\bf p})$ obey  equations $(p_\mu\gamma^\mu-m)u_s({\bf p})=0$ and  $(p_\mu\gamma^\mu+m)v_s({\bf p})=0$, respectively, and $\gamma^{\mu}(\mu=0,1,2,3,4)$ are the gamma matrices.  Equations (\ref{eq1},\ref{eq2}) show that a particle's state $\ket{\psi(t)}$ has a corresponding quantum field state $\ket{\Psi(t)}$, and  this corresponding relation is instantaneous; in some sense the quantum is merely an alternative description of the quantum field.

 We can create a state of one electron with wavefunction $\psi_1(x)=u_s({\bf p})e^{ipx}$, with normalization understood, by acting on the vacuum state $\ket{0}$ with a creation operator  $\ket{{\bf p},s}=b_s^\dagger({\bf p})|0\rangle$. If we does not quantize the field, $b_s({\bf p})$ and $ d_s^\dagger({\bf p})$ in equation \eqref{eq1} are c-numbers, then for the case where there is only one electron, $b_{s'}({\bf p'})=\delta({\bf p}-{\bf p'})\delta_{ss'}$ and $d_{s'}^\dagger({\bf p'})=0$,  equation \eqref{eq1} gives the quantum field $\Psi(x)=u_s({\bf p})e^{ipx}\equiv\Psi_1(x)$. As the excitation of  the quantum field, an arbitrary  electron state $\ket{\psi(x)}$ has a instantaneously corresponding quantum field state $\ket{\Psi(x)}$.

  At the moment when  an electron of state $\ket{\psi}$, which is localized in space, is created, every part of  the  quantum field, which distributes in the whole space, simultaneously makes a corresponding change $\ket{0}\rightarrow\ket{\Psi(x)}$, where $\ket{\Psi(x)}$ is the corresponding quantum field state related to $\ket{\psi}$, which shows that the quantum field is nonlocal, as if the space for the quantum field doesn't exist.  Numerous experimental observations have shown that nonlocality exists among the quantum entangled particles \cite{mard, dmpm, aagp, hzjb,wtbj}, even a single particle can exhibit nonlocality \cite{jdvv,bhpu, gisin}.   It would  be unreasonable if the quantum field itself does not have the character of nonlocality considering that the quantum field is far more fundamental than its excitations and is not localized in space.

What does the quantum state correspond to?   Many physicists have suggested that quantum state does not correspond directly to some physical reality \cite{aepb}, but merely represents an experimenter's knowledge about the reality \cite{leba, cmca, azeg}. Recently Pusey {\it el.} showed a no-go theorem: any model based on both the assumption that the quantum state merely represents knowledge about some physical reality and the assumption that independently prepared systems have independent physical states must make predictions that are incompatible with quantum theory \cite{mpjb}. We notice that $\psi_1(x)=\Psi_1(x)$, which suggests that it is the quantum field that  the wavefunction describes. Thus the quantum state is real and corresponds directly to the quantum field state.

Based on these understandings and the fact that any particle can be considered as the quantum  of the corresponding quantum field we can take a microphysical entity as the excited corresponding quantum field of one quantum. In this model being nonlocal and capable of state superposition  quantum field is responsible for the wave-like nature of the ME, while the quantum is responsible for the ME's particle-like nature.  Here wave-like nature and particle-like nature are no more conflicting, on the contrary they are always harmoniously coexisted  with each being one side of the quantum field.     In each single measurement  detectors are trigged by particles and measured the particle-like properties. While the wave-like properties are shown through the statistical distribution of the data set obtained in numerous repeated measurements, i.e., what the wave-like properties represent is the relationship among the data set.   With this new model many long-standing problems about quantum theory  can now be answered in very natural and simple ways.

 First we discuss the collapse of quantum state on measurement \cite{abgc, aleg, saag}. Consider that the a microphysical entity is in a superposition state $\ket{\psi(x)}=c_1\ket{o_1)}+c_2\ket{o_2)}$, where $c_1$ and $c_2$ are complex numbers satisfying $|c_1|^2+|c_2|^2=1$, $\ket{o_i}(i=1,2)$ are eigenstates of observable $\hat{O}$ with corresponding eigenvalues $o_i$. When a measurement of observable $\hat{O}$ gives a particular result, e.g., $o_1$, the ME 's state is turned into state $\ket{o_1}$ from  $\ket{\psi(x)}$,  the corresponding quantum field makes an instantaneous change $\ket{\psi(x)}\rightarrow\ket{o_1}$. This is the so-called quantum state collapse on measurement. This collapse  is now obvious: the measurement on observable $\hat{O}$ modifies the ME state, resulting in the quantum field state in the whole space making a corresponding changes instantaneously. Because of the nonlocality of the quantum field, this quantum state collapse can take place without difficulty.

 In the particle double-slit experiments the intensity of the particle sources is so low that  while one particle is being recorded, the next one to be recorded is still confined to the particle source, leading to the conclusion that the interference fringe is really collected one by one which shows the particle nature \cite{azei,azeg, lman}. On the other hand the diffraction pattern formed through numerous repeated measurements over a long time  shows the wave nature.  Where does the wave characteristic of the interference come from? Why does a single particle have wave characteristic? From our viewpoint, the particle now is a quantum field of one quantum, and the quantum field has equal field amplitudes at slit $a$ and $b$ for the case of symmetric illumination and is described by $ \ket{\phi}=(\ket{a}+\ket{b})/\sqrt{2}$. The field amplitude at a given observation point ${\bf r}$ is $c({\bf r})=a({\bf r})+b({\bf r})$, and the total probability density  to find the particle at point ${\bf r}$
is given by $p({\bf r})=|c({\bf r})|^2$, where $a({\bf r})$ and $b({\bf r})$ are the amplitudes at point ${\bf r}$  of the corresponding quantum field arrived from slit $a$ and $b$, respectively.  What  the detector recorded in a single measurement is the particle position, the particle-like nature; while the wave-like nature, the diffraction pattern, manifests itself in the distribution of particle positions accumulated in numerous repeated measurements. Thus the wave-like properties of a single particle  originate from  the quantum field. The diffraction pattern requires every single particle simultaneously pass through slits $a$ and $b$, which is unimaginable for some existing interpretations. Here  what pass through the two slits simultaneously is the quantum field, and each particle travels through one slit every time, but which slit the particle passed through is not determined.

When two particles are entangled, e.g., two electron's spins are maximally entangled
\be\label{eq3}
\ket{\psi^+}_{12}=\left(\ket{\uparrow}_1\ket{\downarrow}_2+\ket{\downarrow}_1\ket{\uparrow}_2\right)/\sqrt{2}
\ee
with spin up $\ket{\uparrow}$ and down $\ket{\downarrow}$.
Numerous experiments have shown that there is nonlocal correlation between the entangled electrons.
When a readout of the spin state of electron 1 is performed with a result $\ket{\uparrow}_1$, the spin state of electron 2 is
simultaneously turned into $\ket{\downarrow}_2$,  no matter how far apart they are. There arises a problem: how can this nonlocal correlation take place between two localized particles? We say that the entangled particles interact with each other with the help of the quantum field. Equation \eqref{eq3} shows that the quantum field is in the state where the sum of two electron's spin is zero. When spin 1 is recorded with $\ket{\uparrow}_1$, the quantum field state is simultaneously determined  to be $\ket{\uparrow}_1\ket{\downarrow}_2$, which means that  spin 2 is in state $\ket{\downarrow}_2$. Thus the states of the two electrons' spins are determined at the same time, which is in well agreement with  the recent experimental observations \cite{dsab, ashz, hzjb} and the recent theoretical result that the no-signaling property of quantum correlations excludes any possible explanation of quantum correlations in term of finite-speed influences \cite{jbsp}. The nonlocal quantum field is  a very natural mediator for the interaction between entangled particles.

An arbitrary unknown quantum state can be teleported to one of a pair quantum entangled particles  \cite{dbjp, cbgb}. Why can this teleportation take place considering that there are no direct interaction between these particles?
 We assume that Alice have electrons 1 and 2, and Bob has electron 3. The spin 1  is in an unknown state $\alpha\ket{\uparrow}_1+\beta\ket{\downarrow}_1$ with $|\alpha|^2+|\beta|^2=1$, spins 2,3 are in the maximally entangled state
\be\label{eq4}
\ket{\psi^-}_{23}=\frac{1}{\sqrt{2}}
\left(\ket{\uparrow}_2\ket{\downarrow}_3-\ket{\downarrow}_2\ket{\uparrow}_3\right).
\ee
The state of the considered system has the form
\be\label{eq5}
\ket{\psi}_{123}=(\alpha\ket{\uparrow}_1+\beta\ket{\downarrow}_1)\frac{1}{\sqrt{2}}
\left(\ket{\uparrow}_2\ket{\uparrow}_3-\ket{\downarrow}_2\ket{\downarrow}_3\right).
\ee
This state can be rewritten as
\bea\label{eq9}
&&\ket{\psi}_{123}=\frac{1}{2}[\ket{\psi^-}_{12}(-\alpha\ket{\uparrow}_3-\beta\ket{\downarrow}_3)\notag\\
&&+\ket{\psi^+}_{12}(-\alpha\ket{\uparrow}_3+\beta\ket{\downarrow}_3)
+\ket{\phi^-}_{12}(\alpha\ket{\uparrow}_3+\beta\ket{\downarrow}_3)\notag\\
&&+\ket{\phi^+}_{12}(\alpha\ket{\uparrow}_3-\beta\ket{\downarrow}_3)]
\eea
with the Bell states
\be\label{eq10}
\ket{\psi^\pm}_{12}=\frac{1}{\sqrt{2}}(\ket{\uparrow}_1\ket{\downarrow}_2\pm\ket{\downarrow}_1\ket{\uparrow}_2)
 \ee
 and
 \be\label{eq11}
\ket{\phi^\pm}_{12}=\frac{1}{\sqrt{2}}(\ket{\uparrow}_1\ket{\uparrow}_2\pm\ket{\downarrow}_1\ket{\downarrow}_2).
 \ee
Alice makes a Bell-state measurement on the spins of particles 1 and 2, and finds spins 1 and 2 are in one of four Bell states, e.g., $\ket{\psi^-}_{12}$, leading to the corresponding  quantum field state $
\ket{\psi}^-_{12}(-\alpha\ket{\uparrow}_3-\beta\ket{\downarrow}_3)/\sqrt{2}$,  which shows that spin 3 is in the state $(-\alpha\ket{\bar{\uparrow}}_3-\beta\ket{\bar{\downarrow}}_3)/\sqrt{2}$. The measurement on spins 1 and 2 and the state determination of spin 3 happen simultaneously.  Alice tells Bob his measurement result through classic channel, then Bob carries out a corresponding  unitary transform on particle 3 leaving it in state $(\alpha\ket{\bar{\uparrow}}_3+\beta\ket{\bar{\downarrow}}_3)/\sqrt{2}$. Thus the quantum teleportation of an unknown state is performed with the aid of the quantum field.

Quantum field plays a similar role in Quantum swapping \cite{jpdb, nsbs}. The spins of four electrons  are initialized in the entangled states $\ket{\psi^-}_{12}$ and $\ket{\psi^-}_{34}$, the quantum field is then in the state $\ket{\psi}=\ket{\psi^-}_{12}\ket{\psi^-}_{34}$. This state can be rewritten as follows
\bea\label{eq13}
\ket{\psi}_{1234}&=&\frac{1}{2}[\ket{\psi^+}_{14}\ket{\psi^+}_{23}-\ket{\psi^-}_{14}\ket{\psi^-}_{23}\notag\\
&&-\ket{\phi^+}_{14}\ket{\phi^+}_{23}+\ket{\phi^-}_{14}\ket{\phi^-}_{23}].
\eea
 Alice has electrons 2 and 3, and Bob has electrons 1 and 4. Alice carries out a Bell-state measurement on the spins of two electrons 2 and 3, and find that they are in one of four Bell states, for example $\ket{\psi^+}_{23}$, which shows that  the corresponding  quantum field state is $\ket{\psi}^+_{14}\ket{\psi}^+_{23}$. This quantum field state means that  electrons 1 and 4 are in the state $\ket{\psi^+}_{14}$. The determination of the state of electrons 2 and 3, the quantum field state, and the state of electron 1 and 4 are simultaneous, though Bob doesn't know the state of the electrons at his side until he receives  classic information from Alice.   Thus through the mediation of the nonlocal quantum field quantum entanglement, quantum teleportation, and quantum swapping can be very naturally interpreted.

In the delayed-choice entanglement swapping experiment \cite{xmsz}, two pair of  entangled photons are generated and one photon from each pair is sent to Victor, and two other photons are sent to Alice and Bob, respectively. Alice and Bob detect their photons' polarization states before Victor makes his random choice to project his two photons either onto an entangled state, a Bell-state measurement (BSM),  or onto a separable state, a separable-state measurement (SSM). The Victor's choice of BSM and SSM  projects Alice's and Bob's photons onto an entangled state and onto a separable state, respectively. Victor's delayed choice projects the Alice's and Bob's two already registered photons onto either an entangled state or a separable state, giving rise to the paradox that future actions seem to influence past and already unchangeably recorded events.

 This paradox can be solved as follows: Assuming two entangled photon pairs are in state $\ket{\psi}_{1234}=\ket{\psi^-}_{12}\ket{\psi^-}_{34}$, where $\ket{\ket{\psi^-}_{ij}}=(\ket{H}_i\ket{V}_j-\ket{V}_i\ket{H}_j)/\sqrt{2}$ with the horizontal (vertical) polarization state  $\ket{H}_i (\ket{V}_i)$. Alice and Bob detect the  polarization states of photons 1 and 4  with results, for example $\ket{H}_1\ket{H}_4$, followed by Victor's delayed choice to project photons 2 and 3 onto state $\ket{\phi^+}_{23}$ ( or  $\ket{HH}_{23}$), the quantum field state of four photons is $\ket{\psi_{1234}=\ket{H}_1\ket{H}_4}\ket{\phi^+}_{23}$ (or $\ket{\psi_{1234}=\ket{H}_1\ket{H}_4}\ket{HH}_{23}$ ) with $\ket{\ket{\phi^+}_{23}}=(\ket{H}_2\ket{H}_3-\ket{V}_2\ket{V}_3)/\sqrt{2}$. The victor's delayed choice of BSM doesn't make photons 1 and 4 in entangled state, on the contrary the polarization state of photons 1 and 4 $\ket{H}_1\ket{H}_4$ and the corresponding quantum field state are independent from the Victor's delayed choice. The Victor's delayed choice has no influence on the polarization state of photons 1 and 4. This is the analysis for a single measurement. As for the numerical repeated measurements, Alice's and Bob's data sets can be sorted into several subsets according to Victor's delay choices of measurement and his measurement results. When Victor projected his photons onto entangled state, the distribution of Alice's and Bob's joint data subsets display nonlocal correlation between photons 1 and 4; on the other hand, when Victor projected his photons onto a separable state,  the distribution of Alice's and Bob's joint subsets shows no such nonlocal correlation.

If Victor's  projecting photons 2 and 3 onto $\ket{\phi^+}_{23}$  is  before the measurement on the other two photons with result $\ket{H}_1\ket{H}_{4}$, the quantum field state after a single measurement is $\ket{\psi_{1234}=\ket{H}_1\ket{H}_4}\ket{\phi^+}_{23}$. The quantum field state $\ket{\psi_{1234}=\ket{H}_1\ket{H}_4}\ket{\phi^+}_{23}$ remains the same no matter whether Victor's projection is carried out   before or after  Alice's and Bob's measurements, which is the reason why the statistics of the preselected and postselected data subsets is equivalent \cite{ocoh}. Other delayed choice experiments \cite{ykry, azeg, apps, fktc, cbma, yamz} can be explained in  similar ways.

In summary we have presented a model for a microphysical entity: a ME is taken as the corresponding quantum field which is in the excited state of one quantum. Being the field's quantum an arbitrary  state of the localized quantum  directly corresponds to a quantum field state, different particle states correspond to different quantum field states. This correspondence   shows that the quantum field is nonlocal.  The quantum field is responsible for the wave-like nature of the ME, while the quantum for its particle-like nature. A quantum state corresponds directly to a quantum field state. In a single measurement detectors record the ME's particle-like properties, and its wave-like
properties is shown in the statistical distribution of the data obtained through numerous repeated measurements.
It is the quantum field that makes entangled particles and single particle nonlocal and makes quantum teleportation and quantum entanglement swapping realizable.  With this model the wave/particle dualism, the quantum state collapse on measurement, and the paradox in the delay-choice experiments can be interpreted in a very simple, natural, and reasonable ways.

%\section{CONCLUSIONS}

%\begin{acknowledgments}
 This work was supported by the National Natural Science Foundation of China ( 11072218 ) and by Zhejiang Provincial Natural Science Foundation of China (Grant No. Y6110314).
%\end{acknowledgments}
%%%%%%%%%%%% Bibliography %%%%%%%%%%%%%%%%%%%%%%%%

\end{document}